\documentclass[runningheads]{svmult}

\usepackage{makeidx}   
\usepackage{graphicx}  
\usepackage{subeqnar}  
\usepackage{multicol}  
\usepackage{physprbb}  
\makeindex             

\begin{document}
\title*{Measurement and Information Extraction in Complex 
Dynamics Quantum Computation}
\toctitle{Measurement and Information Extraction in Complex
\protect\newline Dynamics Quantum Computation}
%
%
\titlerunning{Information Extraction in 
Complex Dynamics Quantum Computation}
%
\author{Giulio Casati\inst{1}
\and Simone Montangero\inst{2}}

\authorrunning{Giulio Casati, Simone Montangero}
%
%
\institute{Center for Nonlinear and Complex 
Systems, Universit\`a dell'Insubria and \\
INFM, Unit\`a di Como, Via Valleggio 11, 22100 Como, Italy\\
INFN, Sezione di Milano, Via Celoria 16, 20133 Milano, Italy
\and Scuola Normale Superiore, NEST-INFM,\\
P.zza dei Cavalieri 7, Pisa, Italy\\~\\}

\maketitle              


Quantum Information processing has several different applications: some of 
them can be performed controlling only few qubits simultaneously (e.g. 
quantum teleportation or quantum cryptography) \cite{steane}. 
Usually, the transmission of large amount of information is 
performed repeating several times the scheme implemented for few qubits. 
However, to exploit the advantages of quantum computation, 
the simultaneous control of many qubits is unavoidable \cite{chuang}. 
This situation increases the experimental difficulties of 
quantum computing: maintaining quantum coherence in a large quantum
system is a difficult task. Indeed a quantum computer is a
many-body complex system and decoherence, due to the 
interaction 
with the external world, will eventually corrupt any quantum computation.
Moreover, internal static imperfections can lead to quantum chaos in the
quantum register thus destroying computer operability \cite{GS}. Indeed, 
as it has been shown in  \cite{term},
a critical imperfection strength exists above which the quantum 
register thermalizes and quantum computation becomes impossible.  
We showed such effects on a quantum computer performing an 
efficient algorithm to simulate complex quantum
dynamics \cite{simone1,simone2}.

In this paper, we address a different and very general problem related to the 
extraction of the information. Indeed, the information is 
encoded in the wave function, 
and apart from very particular situations, it is hard to 
find a way to address and extract the useful information. 
Commonly, when trying to extract information, the efficiency of quantum 
information processing is lost. This is one of the main problem to
solve while looking for new quantum algorithms. However, in some
cases, these difficulties can be bypassed, as in Shor, Grover and 
other well known quantum algorithms \cite{ekert}. 

The problem of extracting information is particularly difficult in one of the most
general quantum computing applications: the simulation of many-body complex
quantum systems. Indeed, the results of such simulations is, typically, the quantum
state: the wave function as a whole. The problems is that, in order to measure all $N$ wave function coefficients (coded in
$n_q=log_2(N)$ qubits) by means of a projective measurement, one must repeat the calculus $O(N)$ times. This destroys the efficiency 
of any quantum algorithm even in the case in which such algorithm can compute  the wave function with an exponential gain in the number $n_q$ of elementary gates.

However, as it is the case for other quantum algorithms, there are some questions that can
be answered  
in an efficient way. Here we present an interesting example where important
information can be extracted efficiently by means of quantum
simulations.
We show how this methods work on a dynamical model, the so-called Sawtooth Map \cite{simone1}.
This map is characterized by very different dynamical 
regimes: from near integrable to fully developed chaos; it also exhibits quantum dynamical localization \cite{percival,prosen}. 
We show how to extract efficiently the localization length and the mean square 
deviation.
The results obtained here for the sawtooth map can shed some light for the 
study of different quantum systems. Indeed, as it is
well known, any classical simulation of a quantum system will pretty soon 
be limited by lack of computational resources. 
In our work we show how some questions can be answered efficiently by means 
of a quantum computer, thus allowing the investigation of some general 
properties  beyond the reach of classical supercomputers. 

This paper is organized as follows: in section I we introduce 
our model: the Sawtooth Map. In Section II the quantum
algorithm to compute the quantum motion is presented in detail. In Section III we review 
the exponentially efficient calculation of dynamical localization length. 
Finally, in Section IV, we discuss the additional information 
that can be extracted. Our conclusions are summarized in Section V.

\section{Sawtooth Map}
\label{saw}

\begin{figure}[b]
\begin{center}
\includegraphics[width=.5\textwidth]{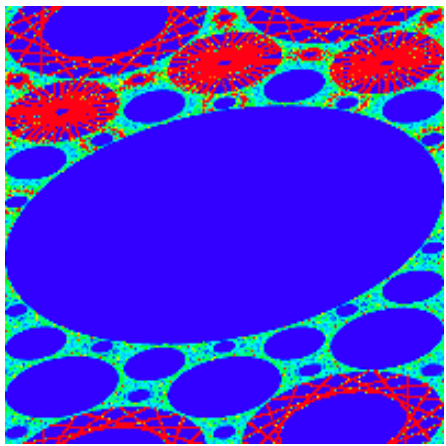}
\end{center}
\caption[]{The phase space of the classical Sawtooth map
  for $K=-0.3$. The phase space density plot is obtained form an ensemble 
of $10^{4}$ trajectories with initial  $n_0$ and random phases outside the main 
island (big blue central area).  $-\pi \le p <\pi$ (vertical axis), 
  $0 \le  \theta < 2 \pi$ (horizontal axis). The probability density is
  proportional to colors: blue for zero, red for maximal
  probability. }
\label{husimi}
\end{figure}

The classical sawtooth map is given by
\begin{equation}
\overline{n}={n}+k(\theta-\pi),
\quad
\overline{\theta}=\theta+T\overline{n},
\label{clmap}
\end{equation}
where $(n,\theta)$ are conjugated action-angle variables
($0\le \theta <2\pi$), and the bars denote the variables
after one map iteration. Introducing the rescaled variable
$p=Tn$, one can see that the classical dynamics depends only on
the single parameter $K=kT$. The map (\ref{clmap}) can be studied on
the cylinder ($p\in (-\infty,+\infty)$), which can also be closed
to form a torus of length $2\pi L$, where $L$ is an integer. 
For any $K>0$, the motion is completely chaotic and one has normal diffusion: 
$<(\Delta p)^2> \approx D(K) t$, where $t$ is the discrete time
measured in units of map iterations and the average $<\cdots>$
is performed over an ensemble of particles with initial momentum
$p_0$ and random phases $0\leq \theta <2\pi$. It is possible to
distinguish two different dynamical regimes \cite{percival}:
for $K>1$, the diffusion coefficient is well approximated by
the random phase approximation, $D(K)\approx (\pi^2/3) K^2$,
while for $0<K<1$ diffusion is slowed down, $D(K)\approx 3.3 K^{5/2}$,
due to the sticking of trajectories close to broken tori
(cantori). For $-4<K<0$ the motion is stable, the phase space
has a complex structure of elliptic islands down to smaller and
smaller scales, and we observed anomalous diffusion,
$<(\Delta p)^2>\propto t^\alpha$, (for example, $\alpha=0.57$ when
$K=-0.1$).  Inside each island the motion can be approximated by an
harmonic oscillator, as can be easily view by the fixed points
stability analysis.  In Fig. \ref{husimi} we show a typical 
phase space of the classical sawtooth map for $K=-0.3$.

The quantum evolution on one map iteration is described
by a unitary operator $\hat{U}$ acting on the wave function
$\psi$:
\begin{equation}
\overline{\psi}=\hat{U}\psi =
e^{-iT\hat{n}^2/2}
e^{ik(\hat{\theta}-\pi)^2/2}\psi,  
\label{qumap}
\end{equation}
where $\hat{n}=-i\partial/\partial\theta$ (we set $\hbar=1$).
Equation (\ref{qumap}) is obtained by integrating over one 
period $T$ the Schr\"odinger equation. As we set $\hbar=1$, one has 
$[\theta,p] = T [\theta,n] = i T$ giving $\hbar_{eff} = T$. Thus,
the classical limit corresponds to $T\to 0$, $k\to \infty$,  and 
$K=kT=\hbox{const}$. In this quantum model one can
observe important physical phenomena like dynamical
localization \cite{prosen,fausto}. Indeed, similarly to other models of quantum chaos \cite{krot},
the quantum interference in the sawtooth map
leads to suppression of classical chaotic diffusion 
after a break time 
\begin{equation} 
t^\star\approx D_n\approx (\pi^2/3) k^2,
\label{loc1}
\end{equation}
where $D_n$ is the classical diffusion coefficient, 
measured in number of levels ($<(\Delta n)^2> \approx D_n t$). 
For $t > t^\star$ only $\Delta n \sim D_n$
levels are populated and the localization length $\ell \sim \Delta n$
for the average probability distribution
is approximately equal \cite{ds1987}:
\begin{equation}
\ell\approx D_n
\label{loc2}
\end{equation} 
Thus the quantum localization can be detected if $\ell$ is smaller 
than the system size $N$.

In Section \ref{secloc}, we study the map (\ref{qumap}) in 
the deep quantum regime of dynamical localization. 
For this purpose, we keep $k,K$ constant. Thus the 
effective Planck constant is fixed and the number 
of cells $L$ grows exponentially with the number 
of qubits ($L=TN/2\pi$). In this case, one studies the 
quantum sawtooth map on the cylinder 
($n\in (-\infty,+\infty)$), which is cut-off 
to a finite number of cells due to the finite 
quantum (or classical) computer memory.

Notice that keeping $K$ and $L$ constant while increasing 
the number of qubits, allows the study of the quantum to classical 
transition. Moreover,  in the stable case, $-4 < K < 0$, one can 
study the wave function evolution inside and outside the islands. 
In the first case, the motion
follows the classical periodic motion with a given characteristic frequency
$\omega$, while outside the classical anomalous diffusion can be suppressed 
by quantum effects. 

We stress again that, since in a quantum computer 
the memory capabilities grow exponentially with 
the number of qubits, already with less than 
40 qubits one could make simulations inaccessible for today's supercomputers.

\section{Quantum algorithm}
\label{alg}
The quantum algorithm introduced in \cite{simone1} simulates 
efficiently the quantum dynamics (\ref{qumap}) using 
a register of $n_q$ qubits. 
It is based on the forward/backward quantum Fourier transform 
\cite{qft} between the $\theta$ and $n$ representations
and has some elements of the quantum algorithm for kicked rotator
\cite{GS1}. 
Such an approach is rather convenient since
the Floquet operator  $\hat{U}$ is the product of two operators 
$\hat U_k = e^{ik(\hat{\theta}-\pi)^2/2}$ and $\hat U_T = e^{-iT\hat{n}^2/2}$:
the first one is diagonal in the $\hat \theta$ representation, 
the latter in the $\hat n $ representation. Moreover both operators
can be decomposed in a sequence of controlled phase shifts without
using any temporary storing register. 

The quantum algorithm for one map iteration requires the following steps:
\\
{\sc I.}  The unitary operator $\hat U_k$ is decomposed 
in $n_q^2$ two-qubit gates 
\begin{equation}
e^{\imath k(\theta -\pi)^2/2} = \prod_{i,j} 
e^{\imath 2 \pi^2 k( \alpha_i 2^{-i} - 
\frac{1}{2n_q})
(\alpha_j 2^{-j} - 
\frac{1}{2n_q})},
\label{dec}
\end{equation}
where $\theta = 2\pi\sum \alpha_i 2^{-i}$, with $ \alpha_i \in \{ 0,1 \} $. 
Each two-qubit gate can be written in the 
$\{|00 \rangle,|01 \rangle,|10 \rangle,|11 \rangle\}$ basis as  
$\exp(i k \pi^2 D)$, where D is a diagonal matrix with elements 
\begin{eqnarray} 
\{
\frac{1}{2 n_q^2}, 
-\frac{1}{n_q}\left(\frac{1}{2^j}-\frac{1}{2n_q}\right) ,
-\frac{1}{n_q}\left(\frac{1}{2^i}-\frac{1}{2n_q}\right) ,\nonumber\\
2\left(\frac{1}{2^i}-\frac{1}{2n_q}\right)
\left(\frac{1}{2^j}-\frac{1}{2n_q}\right)
\}. 
\end{eqnarray} 
{\sc II.} The change from the $\theta$ to the $n$ representation 
is obtained by means of the quantum Fourier transform, which 
requires $n_q$ Hadamard gates and $n_q (n_q - 1)/2$ controlled-phase 
shift gates \cite{qft}.\\
{\sc III.} In the new representation the operator $\hat U_T$ has
essentially the  
same form as $\hat U_k$ in step {\sc I} and therefore it can be decomposed in 
$n_q^2$ gates similarly to equation (\ref{dec}). \\
{\sc IV.}  We go back to the initial $\theta$ representation via inverse 
quantum Fourier transform.\\
On the whole the algorithm requires $3 n_q^2 +n_q$ gates per map iteration.   
Therefore it is exponentially efficient with respect to any known classical 
algorithm. Indeed the most efficient way to simulate the quantum dynamics 
(\ref{qumap}) on a classical computer 
is based on forward/backward fast Fourier transform and requires 
$O(n_q 2^{n_q})$ operations. We stress that this quantum algorithm
does not need any extra work space qubit. 
This is due to the fact that for the quantum
sawtooth map the kick operator $\hat U_k$ has the same quadratic 
form as the free rotation operator $\hat U_T$.

\begin{figure}[b]
\begin{center}
\includegraphics[width=.6\textwidth]{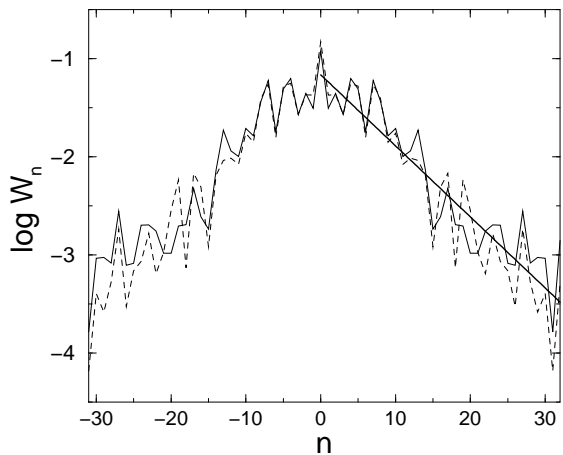}
\end{center}
\caption[]{Probability distribution over the 
momentum basis with $n_q=6$ qubits for $k=\sqrt{3}$ and initial 
momentum $n_0=0$; 
the numerical data are averaged in the intervals $10\leq t \leq 20$
(full curve) and $290\leq t \leq 300$ 
(dashed curve). The straight line fit, 
$W_n\propto \exp(-2|n|/\ell)$, gives the 
localization length $\ell\approx 12$.}
\label{fig2}
\end{figure}

\section{Simulation of dynamical localization}
\label{secloc}


In Fig. \ref{fig2}, we show that, using our quantum algorithm,  
exponential localization can be clearly seen already 
with $n_q=6$ qubits. After the break time $t^\star$, the probability 
distribution over the momentum eigenbasis decays 
exponentially, 
\begin{equation} 
W_n=|\hat{\psi}(n)|^2\approx 
\frac{1}{\ell}\exp\left(-\frac{2 |n-n_0|}{\ell}\right),
\label{expdecay}
\end{equation} 
with $n_0=0$ the initial momentum value. 
Here the localization length is  
$\ell\approx 12$, and classical diffusion 
is suppressed after a break time $t^\star\approx \ell$, 
in agreement with the estimates (\ref{loc1})-(\ref{loc2}). 
This requires a number 
$N_g \approx 3 n_q^2 \ell \sim 10^3$
of one- or two-qubit quantum gates.  
The full curve of Fig. \ref{fig2} shows that an exponentially 
localized distribution indeed appears at $t\approx t^\star$. 
Such  distribution is frozen in time, apart from 
quantum fluctuations, which we partially smooth out by 
averaging over a few map steps. The localization
can be seen by the 
comparison of the probability distributions taken  
immediately after $t^\star$ (full curve in Fig. \ref{fig2}) 
and at a much larger time $t=300\approx 25 t^\star$ 
(dashed curve in the same figure). 

We now discuss how it would be possible to 
extract information (the value of the localization 
length) from a quantum computer simulating the 
above described dynamics. 
The localization length can be measured by 
running the algorithm several times up to 
a time $t>t^\star$. Each run is followed 
by a standard projective measurement on 
the computational (momentum) basis. 
The outcomes of the measurements can be 
stored in histogram bins of width 
$\delta n \propto \ell$, 
and then the localization length can be 
extracted from a fit of the exponential decay 
of this coarse-grained distribution over the 
momentum basis. 
In this way the localization length can be 
obtained with accuracy $\nu$ after the order 
of $1/\nu^2$ computer runs. 
It is important to note that
it is sufficient to perform a coarse grained measurement 
to generate a coarse grained distribution. 
This means that it will be sufficient to measure 
the most significant qubits, and ignore those 
that would give a measurement accuracy below 
the coarse graining $\delta n$. 
Thus, the number of runs and measurements is independent of $\ell$.
However, it is necessary to make about $t^\star \sim \ell$
map iterations to obtain the localized distribution 
(see Eqs. (\ref{loc1},\ref{loc2})).
This is true both 
for the present quantum algorithm and for 
classical computation. This implies that a 
classical computer needs $O(\ell^2\log \ell)$ 
operations to extract the localization 
length, while a quantum computer would 
require $O(\ell (\log \ell)^2)$ elementary gates
(classically one can use a basis size $N \sim \ell$
to detect localization). 
In this sense, for $\ell \sim N=2^{n_q}$ 
the quantum computer gives a 
square root speed up if both classical and quantum computers
perform $O(N)$ map iterations.
However, for a fixed number of iterations $t$
the quantum computation gives an exponential gain.
For $\ell \ll N$ such a gain can be very important
for  more complex physical models, in order to check if the 
system is truly localized \cite{harper}.

\section{Efficient Measurements}
\label{effmis}

We now discuss how some other important system's characteristic quantity 
can be extracted efficiently by means of a quantum simulation. 
We would like to stress that our procedure can be applied 
to any quantum system with a similar dynamics simulated on a quantum computer.

As for the localization length,
also the mean squared moment can be efficiently
extracted with a given precision. Indeed, for a fixed number of
iterations $t$ the quantum computation gives an {\it exponential gain}
since one should compare 
$O(t(\log N)^2)$ gates (quantum computation) with 
$O(tN\log N)$ gates (classical computation).
Starting from the wave function, by repeating $\nu$ projective measurements, 
it is possible to get the mean squared moment and therefore the diffusion coefficient
$D_n\approx <(\Delta n(t))^2>/t$. This is an important 
characteristic which determines the transport properties of the
system.

\begin{figure}[b]
\begin{center}
\includegraphics[width=.4\textwidth]{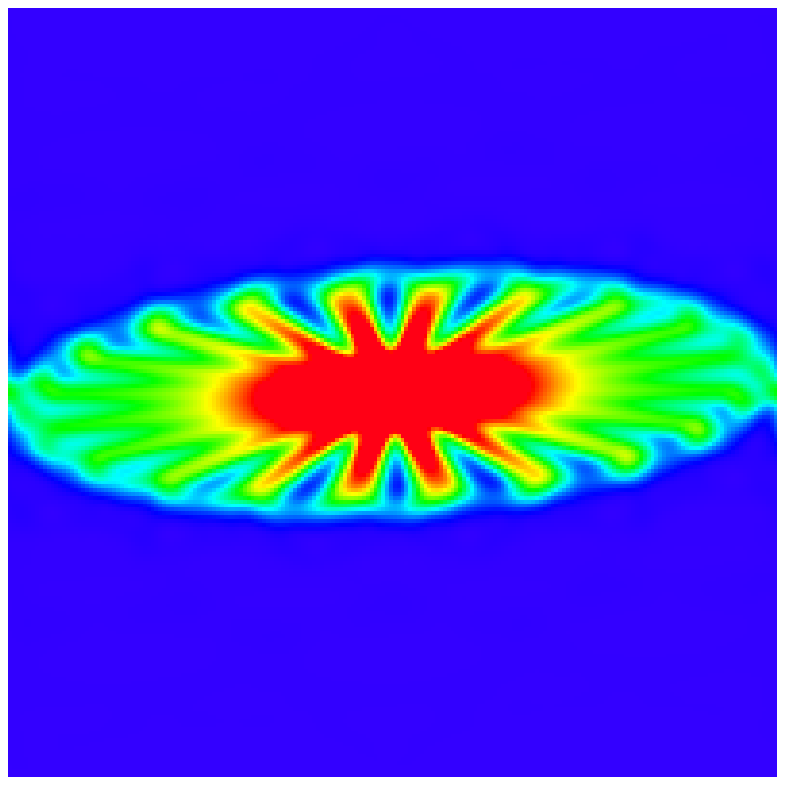} 
\includegraphics[width=.4\textwidth]{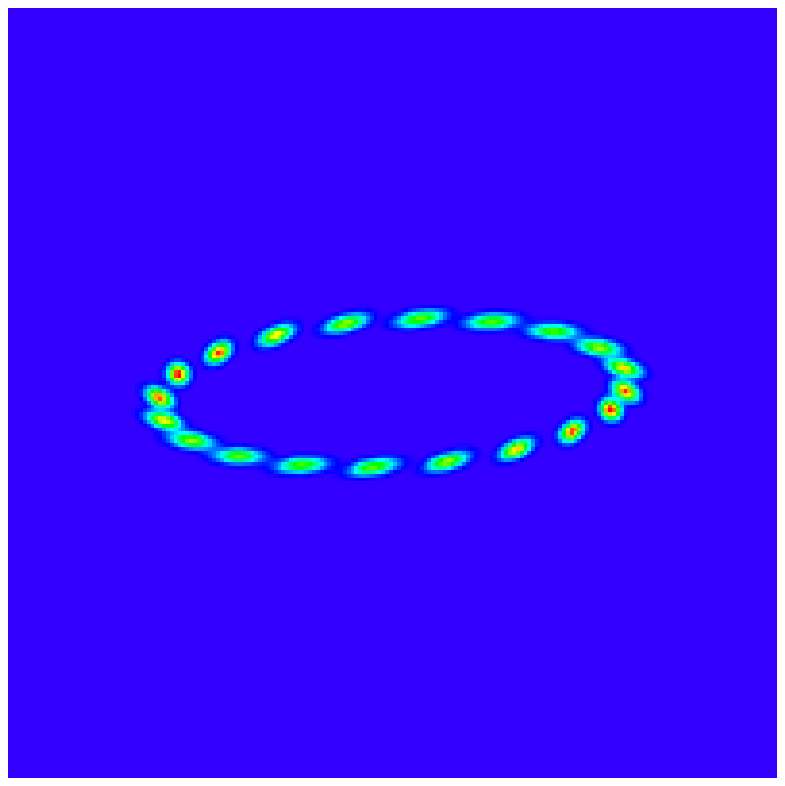}
\end{center}
\caption[]{Husimi function \cite{husimi} of the Sawtooth Map for
  $K=-0.1,T=2\pi/N, s=\Delta p\Delta \theta= 1$, $-\pi \le p <\pi, 
  0 \le  \theta < 2 \pi$. Probability is
  proportional to colors: blue for zero, red for maximal
  probability. Initial conditions: a momentum eigenstate (left), a coherent state (right). The Husimi functions are averaged over twenty 
  map steps.}
\label{figcoe}
\end{figure}

In a similar way one can compute $<(\Delta n(t))^2>$ in the classical limit which is obtained by keeping $K, L$ constant and increasing $N$.
Correspondingly, the number of qubits increases and therefore 
one can explore smaller and smaller scales of the phase space, 
that is, of the wave function. 
This can be fundamental in complex systems with fractal or
self-similar structures in order to determine the nature of the diffusion in
the system, e.g. discriminating from anomalous or brownian diffusion. 
Notice that classically, investigating smaller scales needs exponential efforts. 
On the contrary, only a polynomial 
increasing of resources is needed when performing a quantum simulation. 

Finally we would like to mention another important feature which can be efficiently extracted. 
Indeed if, as in our model for $-4 < K < 0$, an island is present in the system phase space, one can extract the island
characteristic frequency $\omega$. 
Two typical procedures are possible: the first one can be applied if the island
size is comparable with the system size (at least in momentum or phase 
representation). This is the case of the principal island  of our 
model (see Fig. \ref{figcoe}, left).
Indeed, starting from a momentum eigenstate inside the resonance, the
wave function will expand and squeeze periodically. 
Thus, computing the squared moment deviation at different times 
one can recover $\omega$. 
In the case that the island is small compared to system size, one
can start with a localized wave function inside it, e.g. a coherent
state. Although coherent states are more difficult to prepare with respect to 
momentum or phase eigenfunctions, in the case of fixed $K,k$, the
number of operations needed to prepare such states is $N$-independent.
Under the dynamical evolution a coherent state will periodically return back to 
its initial position (Fig. \ref{figcoe}, right). 
As before, measuring the wave
packet center of mass at different times, it is possible to estimate 
the frequency $\omega$.

\section{Conclusions}
We discuss the problem of information extraction in quantum simulation
of complex systems. We present some examples of how quantum
computation can be useful to efficiently extract important information as 
localization length, mean square deviation and system's characteristic 
frequency.
Notice that, in general, extracting the information embedded in the wave function is 
not an efficient process, however we show that in some cases 
quantum simulations can be exponentially efficient with respect to
classical ones. We study in particular the quantum sawtooth map algorithm  which
can be an optimum test for such simulations and requires only a number of qubits less than ten.

Finally, we stress that the procedure discussed here is not restricted to the sawtooth map but can be applied to more general
quantum system. 

This research was supported in part by the EC RTN contract 
HPRN-CT-2000-0156, the NSA and ARDA under 
ARO contracts No. DAAD19-02-1-0086, 
the project EDIQIP of the IST-FET programme of the EC
and the
PRIN-2002 ``Fault tolerance, control and stability in quantum information processing''.


\begin{thebibliography}{99} 

\bibitem{steane} A.Steane, Rep. Prog. Phys. 
{\bf 61}, 117 (1998).
\bibitem{chuang} See, e.g., M.A. Nielsen and I.L. Chuang, 
{\it Quantum Computation and Quantum Information} 
(Cambridge University Press, Cambridge, 2000).
\bibitem{GS} B. Georgeot and D.L. Shepelyansky, 
Phys. Rev. E {\bf 62}, 3504 (2000); {\bf 62}, 6366 (2000)
\bibitem{term} G. Benenti, G. Casati, D. L. Shepelyansky,
Eur. Phys. J. D {\bf 17}, 265 (2001).
\bibitem{simone1} G. Benenti, G. Casati, S. Montangero, and 
D.L. Shepelyansky, Phys. Rev. Lett. {\bf 87}, 227901 (2001).

\bibitem{simone2} G. Benenti, G. Casati, S. Montangero, and 
D.L. Shepelyansky, Eur. Phys. J. D {\bf 20}, 293 (2002);
Eur. Phys. J. D {\bf 22}, 285 (2003).
\bibitem{ekert} A. Ekert, P. Hayden, H. Inamori 
{\it Les Houches Summer School on "Coherent Matter Waves"}, (1999).

\bibitem{percival} I. Dana, N.W. Murray, and I.C. Percival,
Phys. Rev. Lett. {\bf 62}, 233 (1989); Q. Chen, I. Dana,
J.D. Meiss, N.W. Murray, and I.C. Percival, Physica D
{\bf 46}, 217 (1990). 
\bibitem{prosen} G. Casati and T. Prosen, Phys. Rev. E {\bf 59},
R2516 (1999).

\bibitem{fausto} F. Borgonovi, G. Casati, and B. Li, 
Phys. Rev. Lett. {\bf 77}, 4744 (1996); F. Borgonovi, 
Phys. Rev. Lett. {\bf 80}, 4653 (1998); G. Casati 
and T. Prosen, Phys. Rev. E {\bf 59}, R2516 (1999).
\bibitem{geisel} T. Geisel, G. Radons, and J. Rubner,
Phys. Rev. Lett. {\bf 57}, 2883 (1986).
\bibitem{mackay} R.S. MacKay and J.D. Meiss, Phys. Rev. A
{\bf 37}, 4702 (1988).  
\bibitem{prange} R.E. Prange, R. Narevich, and O. Zaitsev,
Phys. Rev. E {\bf 59}, 1694 (1999).
\bibitem{krot} B. Georgeot and D.L. Shepelyansky, Phys. Rev. Lett. 
{\bf 86}, 2890 (2001).
\bibitem{ds1987} D.L. Shepelyansky, Physica D {\bf 28}, 103 (1987).
\bibitem{qft} See, e.g., A. Ekert and R. Jozsa, Rev. Mod. Phys.
{\bf 68}, 733 (1996).
\bibitem{GS1} B. Georgeot , and D.L. Shepelyansky, Phys. Rev. Lett.
{\bf 86}, 2890  (2001).
\bibitem{harper} T. Prosen, 
I.I. Satija, and N. Shah, Phys. Rev. Lett. {\bf 87},
066601 (2001), and references therein. 
\bibitem{husimi} The computation of Husimi functions is 
described in S.-J. Chang and K.-J. Shi, Phys. Rev. A 
{\bf 34}, 7 (1986).

\end{thebibliography}
\end{document}